# High Efficiency Power Side-Channel Attack Immunity using Noise Injection in Attenuated Signature Domain

Debayan Das[1], Shovan Maity[1], Saad Bin Nasir[2], Santosh Ghosh[3], Arijit Raychowdhury[2], Shreyas Sen[1]

[1]School of Electrical and Computer Engineering, Purdue University, USA
[2]School of Electrical and Computer Engineering, Georgia Institute of Technology, USA
[3]Intel Labs, Hillsboro, Oregon, USA
{das60, shreyas}@purdue.edu

*Abstract*— With the advancement of technology in the last few decades, leading to the widespread availability of miniaturized sensors and internet-connected things (IoT), security of electronic devices has become a top priority. Side-channel attack (SCA) is one of the prominent methods to break the security of an encryption system by exploiting the information leaked from the physical devices. Correlational power attack (CPA) is an efficient power side-channel attack technique, which analyses the correlation between the estimated and measured supply current traces to extract the secret key. The existing countermeasures to the power attacks are mainly based on reducing the SNR of the leaked data, or introducing large overhead using techniques like power balancing. This paper presents an *attenuated signature AES* (AS-AES), which resists SCA with minimal noise current overhead. AS-AES uses a shunt low-drop-out (LDO) regulator to suppress the AES current signature by 400× in the supply current traces. The shunt LDO has been fabricated and validated in 130 nm CMOS technology. System-level implementation of the AS-AES along with noise injection, shows that the system remains secure even after $50K$ encryptions, with $10\times$ reduction in power overhead compared to that of noise addition alone.

*Keywords*— Side Channel Attack (SCA), Cryptographic Hardware, Power Analysis Attack, Countermeasure, Attenuated Signature AES, Shunt LDO, Noise Injection.

I. INTRODUCTION

The rapid proliferation of mobile devices and wireless communications has enhanced the need for stronger encryption algorithms. However, hardware implementations of these computationally-secure cryptographic algorithms provide sufficient *"side-channel"* information for the attackers to decipher the secret key. Side-channels include power consumption [1], [2], electromagnetic emanations [3], [4], acoustic vibrations [5], or the timing of various encryption operations [6], [7].

*A. Motivation*

For an attacker, the power monitoring attack is one of the most common side-channel attacks on modern computer systems. The attacker measures the power consumption of the encryption device under test (DUT), and performs subsequent simple (SPA) [8] or differential power analysis (DPA) [1] of the obtained traces to decipher the secret key.

In this article, we focus on the correlational power analysis (CPA) attack [9] on a 128-bit Advanced Encryption Standard (AES) engine. However, the proposed AS-AES is a generic technique to resist power attacks, and can be applied to other encryption algorithms. Figure 1 gives an overview of the proposed signature attenuating AES (AS-AES). The underlying idea is to embed the AES in a signature attenuating hardware, such that the variations in the AES current is highly suppressed and is not reflected in the supply current traces, thereby requiring much lower noise current injection to decorrelate the measured supply traces.

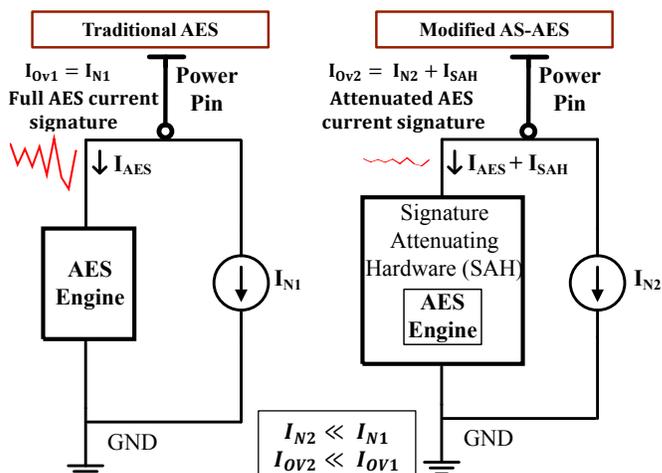

Figure 1: Traditional AES and Proposed AS-AES with noise injection: A comparative Overview ($I_{Ov}$: overhead current)

*B. Contribution*

This paper proposes a new hardware-based technique for cryptographic devices to resist power analysis attacks (PAA). Specific contributions of this paper are:

- Demonstrated the relation between only noise injection vs. correlation (for CPA with correct key), both mathematically and through simulation results. This highlighted the inefficiency of only noise injection leading to a 370% power overhead requirement for achieving Minimum Traces to Disclosure (MTD)$> 50K$.
- Proposed a new generic DPA/CPA countermeasure that can be applied to any cryptographic algorithm. This is achieved by embedding the cryptographic engine within a high-efficiency signature attenuation hardware, to reduce ($> 400\times$) the secret signature on power pin. Injection of noise to mask SCA in the Attenuated-Signature AES (AS-AES) domain, reduces noise-injection overhead (~1 $mA$) for SCA immunity by $70\times$, compared to standalone noise injection.
- Demonstrated that with the proposed AS-AES, none of its secret key blocks have been disclosed ($MTD > 50K$) with

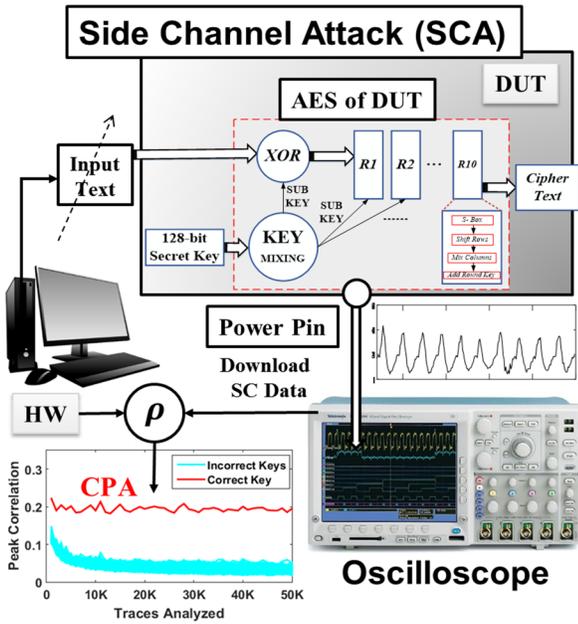

Figure 2: Power Side-Channel attack on unprotected 128-bit AES

> 10× lower power overhead compared to parallel noise incorporation.

## II. RELATED WORK

Immunity to power analysis attack could be improved by modifying the encryption hardware module such that the power consumption of the device is independent of the signal transitions. The existing protection schemes for the hardware encryption engines against power analysis attacks can be classified into three major categories. The first category reduces the SNR of the leaked information by injecting noise into the supply traces [10]. However, solely noise addition is not an optimum technique to make a power attack infeasible. A statistical analysis for this technique is presented in Section III. Also, the effect of power delivery network (PDN) on SNR has been studied and a frequency-dependent noise injection circuit was implemented by Wang et al. [11].

The second category reduces side-channel leakage by balancing the power consumption of the rising and falling transitions. Power balancing logic implementations include the sense-amplifier based logic (SABL) [12], dual-rail circuits [13], [14], and wave dynamic differential logic (WDDL) [15]. The WDDL structure appears to be the first power attack resistant power-balancing circuit validated in silicon with a *Measurements to Disclosure* (MTD) of *~21K*. However, the increased protection consumed 4× power overhead, 3× area, and a 4× performance degradation.

The third category of protection reduces the power side-channel leakage information by isolating the supply from the encryption engine. It includes switched capacitor techniques [16], [17], [18] and integrated voltage regulator (IVR) based implementations [19], [20], [21], [22]. The switched capacitor current equalizer module proposed by Tokunaga et al. [17] demonstrated Power SCA immunity, however, it resulted in 2× performance degradation and 33% power overhead. The impact of package parasitic and integrated buck converters on power SCA vulnerability have been analyzed in [11], [19]. Another concept is to degrade the performance of IVR such that supply current has lesser correlation with the AES current. Implementations of this concept could be found using Analog LDO [20], Digital LDO (resolution is traded off, droop increases) [21] and a buck converter [22]. The above described techniques exhibit a fundamental trade-off between system performance (e.g. dynamic loop response) and reduction in side-channel vulnerability.

In this work, the proposed AS-AES effectively combines the first and third categories of protection to achieve SCA immunity with high power efficiency and no performance degradation. Figure 2 shows a block diagram representing the CPA attack on the AES engine of the device under attack (DUT). The traditional AES-128 core revealed one byte (1st byte in this example) of the secret key with only $1K$ traces, while the same attack applied to the modified AS-AES core could not extract the key byte even with $50K$ traces.

## III. POWER SIDE-CHANNEL IMMUNITY USING NOISE INSERTION

Noise injection into power consumption measurements is a convenient approach to defend against power side-channel attacks [23]. The correlation ($\rho_{TH}$) between the estimated hamming weight matrix *(H)* and the obtained power traces *(T)* can be given as,

$$\rho_{TH} = \frac{Cov\,(T,H)}{\sigma_T * \sigma_H}$$

where, *Cov* denotes the covariance matrix, and $\sigma_T, \sigma_H$ represents the standard deviation of *T and H* respectively. Now, the main goal is to add enough random noise to resist a side-channel attack, yet introducing a minimal power overhead in the system. If *N* denotes the amount of random noise added into the circuit, and $T' = T + N$ denote the modified traces, the modified correlation factor $\rho_{T'H}$ can be written as,

$$\rho_{T'H} = \frac{Cov(T+N,H)}{\sigma_{T+N} * \sigma_H}$$
$$= \frac{E[(T+N)*H] - E(T+N)E(H)}{\sqrt{E(T+N)^2 - E^2(T+N)} * \sigma_H}$$
$$= \frac{E[T*H] - E(T)E(H)}{\sqrt{E(T)^2 + E(N)^2 - E^2(T) - E^2(N)} * \sigma_H}$$
$$= \frac{Cov(T,H)}{\sqrt{\sigma_T^2 + \sigma_N^2} * \sigma_H}$$

*E* represents the expectation or mean of the corresponding variable. The independence of the variables *N* and *H* has been utilized in this expression. Thus, we see that, higher is its variance $\sigma_N^2$, and lower is the correlation $\rho_{T'H}$. However, noise insertion alone introduces a significant current overhead.

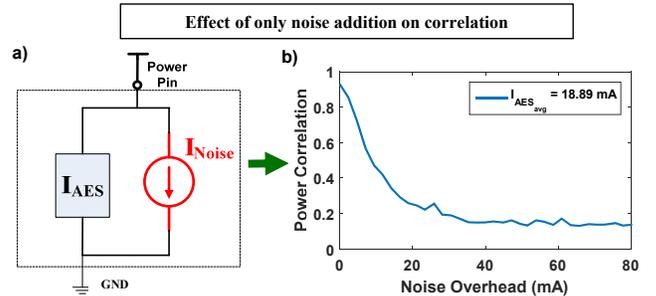

Figure 3(a, b): Effect of Noise insertion alone, on current overhead for power SCA immunity

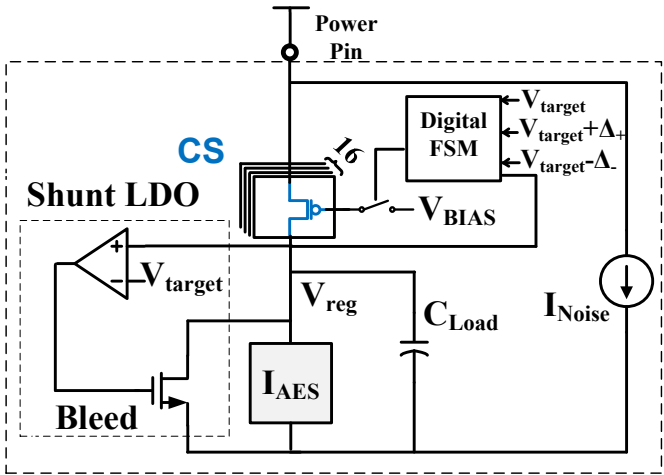

Figure 4: Proposed AS-AES architecture with noise injection to defend against power side-channel attacks

Figure 3 shows the effect of noise insertion on power attack (PAA) immunity. It can be seen that for the actual key, the traces correlate in absence of noise ($\rho_{T'H} \sim 0.9$). For the AES-128 core under attack, the average current consumption of the sampled traces during the AES operations was found to be $\sim 18.89\ mA$. In order to provide high enough resistance to a DPA or CPA, noise insertion requires a current overhead of $\sim 70\ mA$ (refer MTD plots in Figure 12 (a-d)), which incurs a current overhead of approximately *four times* (370%) of the average AES current consumption. Thus, only noise insertion is not an optimized solution for power SCA immunity.

## IV. PROPOSED ATTENUATED SIGNATURE AES (AS-AES)

In this section, we introduce the *attenuated signature AES* (AS-AES) with noise injection. The proposed AS-AES architecture (Figure 4) incorporates a shunt LDO, and an integrating load capacitor. As discussed earlier, to remove the correlation spikes for the actual key, the supply current needs to be independent of the AES current. As seen from Figure 4, the voltage across AES block ($V_{reg}$) needs to be regulated at a constant value, with minimum power overhead. Traditional LDOs, which are commonplace in ICs, aim to maintain a constant $V_{reg}$, however, the supply current reflects the changes in load (e.g. AES) current. In our design, the goal is to maintain a constant $V_{reg}$, as well as to make the supply current independent of the AES current during encryption operations.

### A. Operation Principle

The proposed AS-AES operates using two control loops. Switched-mode control (SMC) is implemented using a low bandwidth slow loop to track large changes in the average AES current. The concept of SMC has been demonstrated in [24]. The SMC loop (Figure 5) regulates the current through an array of PMOS current sources, which are activated when the output voltage goes beyond $V_{target} \pm \Delta$. Here $\Delta$ acts as a guard-band and prevents the digitally controlled loop from being continuously ON. As soon as the output voltage is within $\Delta$, the digital loop is gated and the analog shunt regulator takes the output voltage to $V_{reg}$. For the AES-core under attack, it was verified that the average current ($\sim 18.89\ mA$) consumed by the AES engine for different inputs remain almost the same throughout the encryption operations, as it performs repetitive ten S-box operations for each input byte. SMC only engages to compensate for slow changes in the average AES current. Due to the almost-constant nature of the average AES current, the SMC loop stays disengaged and the PMOS (Figure 4) acts as a current source (CS), in saturation, with high drain to source impedance ($r_{ds}$).

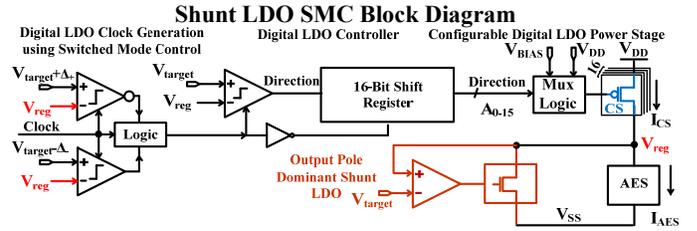

Figure 5: Block diagram of the SMC loop with shunt LDO in AS-AES

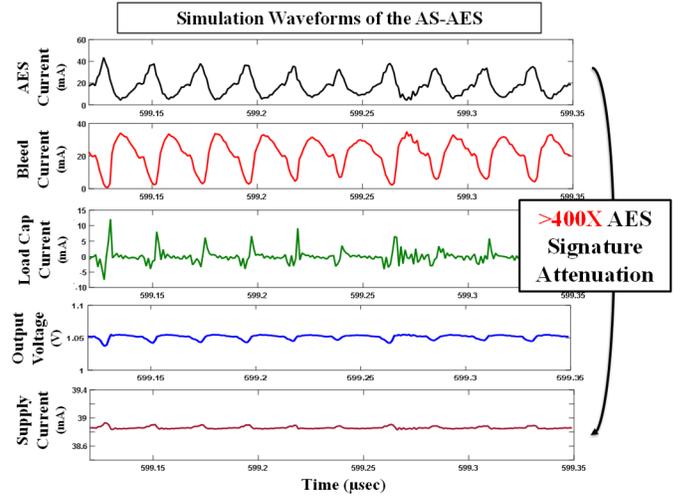

Figure 6: Snapshot of the time-domain waveforms of the AS-AES

Once the SMC loop is set, the CS current is fixed at a value higher ($I_{CS} = I_{AES_{avg}} + I_{ov}$) than the average load (AES current). The fast loop incorporates a shunt LDO with a NMOS bleed to sink any excess current from the supply, when the AES current consumption is lower than the average. Thus, the bleed restricts unnecessary charging of the load capacitor ($C_{Load}$), and thereby regulates the output voltage ($V_{reg}$). On the other hand, when the AES current requirement is more than the supply current, the capacitor provides the necessary extra current, thereby maintaining a constant supply current irrespective of the AES current variation, at the expense of an instantaneous droop in $V_{reg}$, which is reduced with increased $C_{Load}$ or $I_{ov}$.

### B. Steady State Analysis of the AS-AES

The parameters involved in the design of the fast control loop (steady state) of the AS-AES are the NMOS bleed size, the choice of integrating load capacitor, gain ($A_v$) and bandwidth of the operational amplifier (OP-AMP) in the shunt-LDO loop (Figure 4). When the transistors are properly biased into saturation and considering steady state operation, the small-signal analysis of the AS-AES can be performed (Figure 7(a)). The small-signal analysis is shown as follows:

$$-g_{ds}v_{reg} = i_{AES} + v_{reg}SC_{Load} + \frac{aG_m}{S+a}v_{reg}; \quad G_m = A_v g_m.$$

$$\frac{v_{reg}}{i_{AES}} = -\frac{1}{g_{ds} + SC_{Load} + \frac{aG_m}{S+a}}$$

$$\frac{i_{CS}}{i_{AES}} = \frac{g_{ds}}{g_{ds} + SC_{Load} + \frac{aG_m}{S+a}}$$

$$AF = \frac{g_{ds}}{C_{Load}} \frac{s+a}{\left[S^2 + S\left(a + g_{ds}/C_{Load}\right) + (aG_m + ag_{ds})/C_{Load}\right]}$$

where, AF is the AES signature attenuation factor; $g_m$ and $a$ represents the transconductance and the dominant pole of the OP-AMP, respectively. Figure 8 shows the magnitude Bode plot of the attenuation factor (AF) with nominal design, along with its effect on different key parameters. AF increases as loop gain ($G_m$) is increased. Increasing $a$ (i.e. shunt LDO loop bandwidth) increases the rejection range. Higher $C_{load}$ helps with high frequency rejection. In general, the important point to note from these analysis is that for all cases, AS-AES provides high attenuation of AES signature current to supply current.

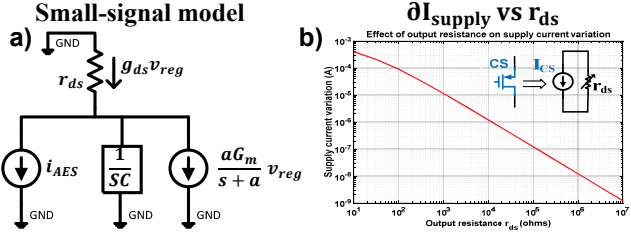

Figure 7: (a) Small-signal model of the AS-AES; (b) Effect of output resistance of the current source on supply current variation (Log-Log scale)

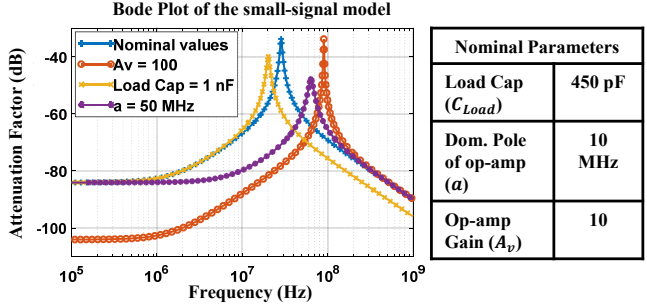

Figure 8: Magnitude Bode plot for attenuation factor (AF) of the AS-AES

Ideally, if the integrating capacitor is high enough, so as to deliver any excess current drawn by the AES, and if the bleed transistor is strong enough to sink any extra current, and the shunt LDO loop bandwidth is very high, the droop in the output voltage $V_{reg}$ would be negligible. However, bandwidth of the fast loop is limited due to the presence of a non-dominant pole at the gate of the bleed NMOS. Also, the choice of capacitor has a trade-off with the area. The choice of the optimized design parameters are discussed in Section V.

Figure 6 shows the time-domain waveforms of the AS-AES during an encryption operation. It can be seen that when the AES current goes low, the bleed sinks the excess current, not allowing the load capacitor to charge up, and thereby maintaining a constant $V_{reg}$. Output voltage ($V_{reg}$) with low ripples ensure that the AES current variation is sufficiently suppressed in the supply current. When the AES current is high, bleed provides the current until it runs out range. When the bleed device turns off, the load capacitor delivers the required current, thereby maintaining the supply current at a constant value, however causing a small droop. In Figure 6, an average droop of ~10 $mV$ in $V_{reg}$ can be observed whenever the AES current shows a rising spike, using a 450 pF integrating capacitor. High output resistance ($r_{ds}$) of the PMOS minimizes supply current variations caused by the signature of the small $V_{reg}$ variations. Since an ideal constant current source (with infinite impedance) is not practical, a finite $r_{ds}$ would still reflect the relative change in voltage $V_{reg}$ in the supply current, however, it will be highly attenuated. AS-AES achieves the AES current signature attenuation of >400× in the supply current.

*C. Noise Injection on AS-AES*

The peak-to-peak variation in the supply current with different $r_{ds}$ of the current source PMOS is shown in Figure 7(b). Even designing a current source with output resistance in the order of $M\Omega$ would reflect a variation in the order of $nA$, which can still be measured using an oscilloscope. Thus, a change in AES current in the order of $mA$ gets reflected in the supply current in an attenuated scale of the order few $\mu A$, depending on the impedance of the current source. Although the signature gets attenuated, the change in current will still be reflected in the power supply traces, and it cannot yet resist a CPA attack completely, resulting in some correlational peaks.

Note, though the signature is still present it is highly attenuated. A small amount of random noise current is injected (as shown in Figure 4) in order to decorrelate the traces with the estimated hamming weight matrix, and thereby provide significant immunity against CPA attack. The noise current generation circuit is a simple linear feedback shift register

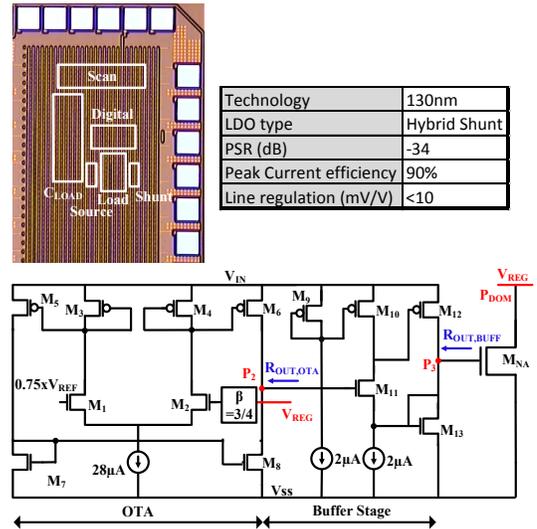

Figure 9: Proof-of-concept shunt LDO design measured on a 130nm testchip. Die-photo, key measurement results and the shunt LDO circuit is shown

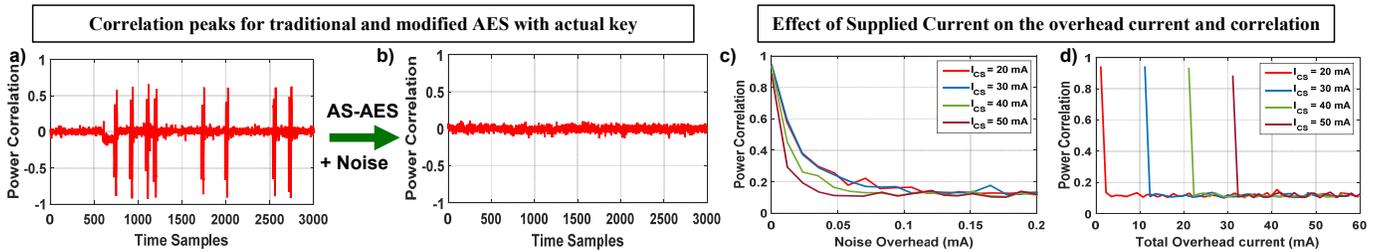

Figure 10: (a, b): CPA attack on traditional and modified AES (with noise addition) respectively, for the actual key; (c, d): Effect of PMOS source current on the overhead current required to provide immunity against power SCA.

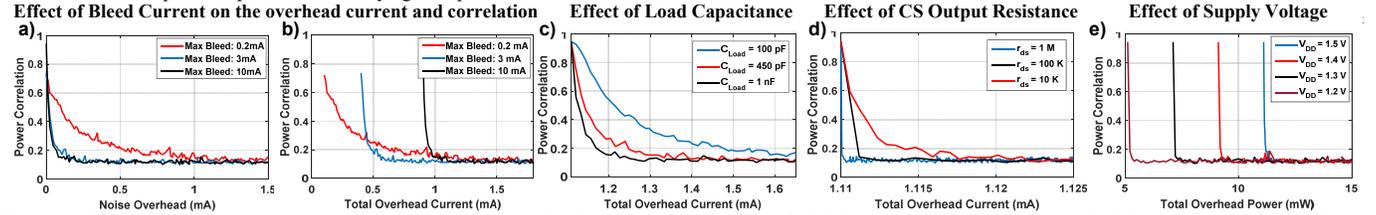

Figure 11: Effect of design parameters on the noise and total overhead current/power required for high power attack immunity. (a, b): Bleed current sink capacity; (c) Load capacitance; (d) PMOS current source output resistance ($r_{ds}$); (e) Supply voltage ($V_{DD}$)

(LFSR) with a current-steering DAC. This noise overhead is very minimal compared to the 'only noise' addition technique, and depends on the output resistance of the current source PMOS, and small signal variation present in $V_{reg}$. Thus, the total current overhead includes the current consumed by the shunt LDO and the noise overhead.

### D. Shunt LDO Design and Validation

The feasibility of a linear regulator featuring SMC and shunt regulation has been shown in Figure 9. A proof-of-concept design has been fabricated in the GF 130 nm CMOS technology. The digital control loop allows large signal control of the input current and is activated when the output voltage goes beyond $V_{target} \pm \Delta$. The shunt analog regulator is a two stage design with an operational transconductance amplifier (OTA) based first stage and a shunt feedback buffer based second stage. In the current test-chip, Δ=70mV has been chosen. The shunt (bleed) path is designed to consume 5% to 40% of the load (AES) current and provide regulation. The phase margin in the worst case load condition is 88°. A summary of the measurements on the test-chip is also shown. Peak current efficiency of 90% is measured. This measured line regulation is less than $10\ mV/V$. This design illustrates the concept of an SMC based shunt LDO and the capability of the design to provide regulation across a load range.

## V. RESULTS

### A. CPA attack on the AES-128 core

We perform CPA attack on the AES-128 power traces. Figure 10(a) shows the correlational peaks for the actual key when the unprotected AES was subjected to CPA attack (< 1K traces). The same attack on modified AES (AS-AES) does not reveal the secret key, highlighting the SCA immunity (Figure 10(b)).

### B. Design Space Exploration

Figure 10(c,d) and Figure 11(a-d) show the choice of parameters for the AS-AES design. Figure 10(c) shows that, as the supply current ($I_{CS}$) is increased (i.e. the overhead/bleed current is increased), the required noise current (overhead) reduces. However, $I_{CS}$ of $20\ mA$ is sufficient for $I_{AES_{avg}}$ of $18.89\ mA$ (Figure 10(d)), as it reduces correlation significantly with minimum current overhead. Now, $I_{ov} = I_{bleed} + I_{noise} + I_{opamp}$. Considering $I_{CS} = 20\ mA$, $I_{bleed} = I_{CS} - I_{AES_{avg}} \sim 1.1\ mA$. Hence, the bleed should be designed to have a minimum current sink capacity of $1.1\ mA$. However, higher bleed capacity increases the current overhead, and hence should be optimized. Figure 11(a, b) justifies that the bleed current sink capacity of $\leq 3\ mA$ would be an optimal choice in terms of the noise injection and $I_{ov}$, since higher bleed would not reduce the correlation any further, but increase the $I_{ov}$, which is not desired. Figure 11(c) shows that the overhead current can be lowered with higher load capacitance. However, considering its trade-off with area and the current overhead, $450\ pF$ load capacitance is chosen over $1\ nF$. Figure 11(d) shows that higher output resistance ($r_{ds} = 1\ M\Omega$) of the PMOS current source allows lower noise current overhead and thus the total overhead current is reduced. Figure 11(e) shows that increasing the supply $V_{DD}$ (i.e. the dropout in the LDO, increasing $r_{ds}$) increases the power overhead ($P_{ov}$), as $V_{reg}$ is maintained at a particular voltage ($1\ V$ in our case).

The above results were obtained with 1000 traces, and gives a good sense of the choice of design parameters and its effects. To confirm the exact overheads and efficiency required, the CPA attack is run over significantly more traces. As the number of traces analyzed is increased, the correlation coefficients ($\rho$) are expected to be more accurate, and $\rho$ for the correct key should be consistently higher than the $\rho$ of the peak (absolute maximum) correlation of the incorrect keys.

### C. SCA Immunity Verification with MTD Analysis

Figure 12(a-d) (noise addition alone) and Figure 12(e-h) (AS-AES + noise injection) shows the evolution of the correlation coefficient as the number of encryption run increases for different noise injection. Figure 12 (a-d) shows that noise addition alone, involves an overhead of $70\ mA$ to achieve Measurements to Disclosure (MTD) >50$K$. As the AS-AES core is subjected to a CPA attack, it is seen that MTD

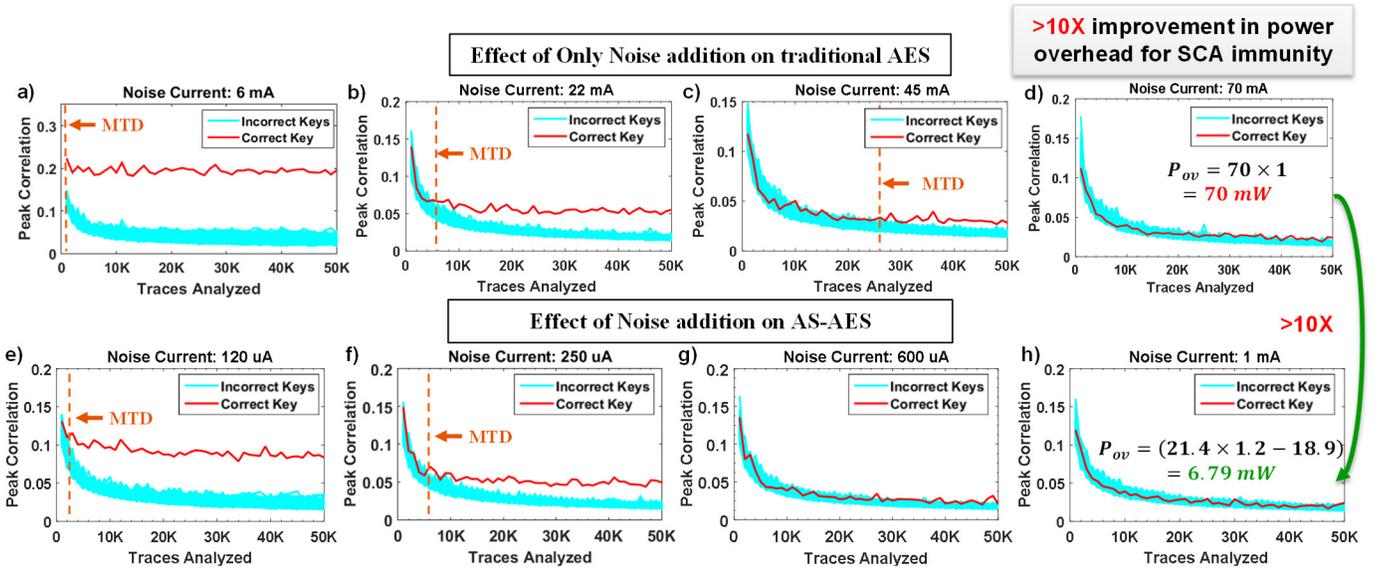

Figure 12: (a-d) MTD plots for only noise addition on traditional AES; (e-h): MTD plots for noise injection on AS-AES.

$>50K$ is achieved, with only a noise current injection of $1\ mA$ (Figure 12(h)). Also, the current consumed by the OP-AMP is $400\ \mu A$. Hence, the total overhead current (as discussed previously) is $I_{ov} = I_{bleed} + I_{noise} + I_{opamp} = 1.11\ mA + 1\ mA + 0.4\ mA = 2.51\ mA$, which is 28× lower than that of noise addition alone. The total overhead power for the AS-AES architecture is $(20\ mA + 1\ mA + 0.4\ mA) * 1.2\ V - 18.89\ mA * 1\ V = 6.79\ mW$. In case of solely noise insertion the power overhead required is $P_{ov} = 70\ mA * 1\ V = 70\ mW$. Thus, AS-AES along with noise injection provides $>10×$ improvement in the power overhead for iso-SCA immunity, compared to only noise addition, with $MTD > 50K$. Power efficiency for AS-AES is given as, $\eta = \frac{18.89\ mA * 1\ V}{21.4\ mA * 1.2\ V} * 100 = 73.56\ \%$ (includes noise overhead). Hence, AS-AES consumes similar overhead as [17], but does not incur the performance penalty.

## VI. CONCLUSION

Power Side Channel Attack is a prominent attack on cryptographic ICs. This work proposes low-overhead hardware modification that attenuates critical AES signature (AS) by $>400×$ in the supply current which the attackers could observe from the periphery of an encryption ASIC. Noise injection in the AS domain achieves SCA immunity with extremely high-efficiency. Successful SCA immunity is demonstrated for a CPA side-channel attack, with up to $50\ K$ traces. Power SCA immunity is achieved with 73.56 % power efficiency, and 10× lower power overhead, compared to only noise addition and more importantly without imposing any performance penalty.

## REFERENCES

[1] P. C. Kocher et. al. "Differential Power Analysis," in *Proceedings of the 19th CRYPTO* London 1999, pp. 388–397.
[2] E. Brier et.al. "Optimal Statistical Power Analysis," 152, 2003.
[3] J.-J. Quisquater et.al. "ElectroMagnetic Analysis (EMA): Measures and Counter-measures for Smart Cards," in *SpringerLink*, Springer Berlin Heidelberg, pp. 200–210.
[4] K. Gandolfi, et.al. "Electromagnetic Analysis: Concrete Results," in *CHES*, London, UK, UK, 2001, pp. 251–261.
[5] D. Genkin, et.al. "RSA Key Extraction via Low-Bandwidth Acoustic Cryptanalysis," in *SpringerLink*, Springer Berlin Heidelberg,
[6] D. Brumley et.al. "Remote timing attacks are practical," *Comput. Netw.*, vol. 5, no. 48, pp. 701–716, 2005.
[7] P. C. Kocher, "Timing Attacks on Implementations of Diffie-Hellman, RSA, DSS, and Other Systems," in *CRYPTO* London, UK, 1996,
[8] C. Clavier, et.al. "Simple Power Analysis on AES Key Expansion Revisited," in *SpringerLink*, Springer Berlin Heidelberg, pp. 279–297.
[9] E. Brier et.al. "Correlation Power Analysis with a Leakage Model," in *Cryptographic Hardware and Embedded Systems - CHES 2004*,
[10] T. Güneysu et.al. "Generic Side-Channel Countermeasures for Reconfigurable Devices," in *SpringerLink*, Springer Berlin Heidelberg,
[11] X. Wang *et al.*, "Role of power grid in side channel attack and power-grid-aware secure design," in *50th (DAC)*, 2013, pp. 1–9.
[12] K. Tiri et.al. "A dynamic and differential CMOS logic with signal independent power consumption to withstand differential power analysis on smart cards," in *ESSCIRC 2002*. pp. 403–406.
[13] M. Bucci et.al. "Three-Phase Dual-Rail Pre-Charge Logic," in *SpringerLink*, Springer Berlin Heidelberg, pp. 232–241.
[14] D. Sokolov et.al."Design and analysis of dual-rail circuits for security applications," *IEEE Trans. Comp.*, vol.54, no.4, pp.449–460
[15] D. D. Hwang *et al.*, "AES-Based Security Coprocessor IC in 0.18- CMOS with Resistance to Differential Power Analysis Side-Channel Attacks", *JSSC*, vol.41, no.4, pp.781–792, Apr. 2006.
[16] A. Shamir, "Protecting Smart Cards from Passive Power Analysis with Detached Power Supplies," in *SpringerLink*, Springer Berlin Heidelberg,
[17] C. Tokunaga et.al. "Securing Encryption Systems with a Switched Capacitor Current Equalizer," *IEEE JSSC*, vol. 45, no. 1, Jan. 2010.
[18] P. Corsonello et.al. "An integrated countermeasure against differential power analysis for secure smart-cards," in *IEEE ISCAS*, 2006
[19] M. Kar et.al. "Impact of inductive integrated voltage regulator on the power attack vulnerability of encryption engines: A simulation study," in *CICC* 2014, pp. 1–4.
[20] A. Singh et.al. "Exploring power attack protection of resource constrained encryption engines using integrated low-drop-out regulators," in *ISLPED*, 2015, pp. 134–139.
[21] A. Singh et al. "Integrated all-digital low-dropout regulator as a countermeasure to power attack in encryption engines," *HOST*, 2016
[22] M. Kar, et.al. "Exploiting Fully Integrated Inductive Voltage Regulators to Improve Side Channel Resistance of Encryption Engines," in *ISLPED 2016* New York, USA, 2016, pp. 130–135.
[23] T. S. Messerges et.al. "Examining Smart-Card Security Under the Threat of Power Analysis Attacks," *IEEE Trans Comp.*, vol. 51, no.5, May 2002.
[24] S. B. Nasir et.al. "A 130nm hybrid low dropout regulator based on switched mode control for digital load circuits," in *ESSCIRC Conference* 2016, pp. 317–320.